%% file: main.tex
\documentclass[sigconf,natbib=true]{acmart}
\AtBeginDocument{%
  \providecommand\BibTeX{{%
    \normalfont B\kern-0.5em{\scshape i\kern-0.25em b}\kern-0.8em\TeX}}}

\newcommand{\fixme}[1]{\textcolor{black}{#1}}
\newcommand{\bheading}[1]{\vspace*{.5em}\noindent{\textbf{#1.}}}

\copyrightyear{2023}
\acmYear{2023}
\setcopyright{acmlicensed}\acmConference[COMPASS '23]{ACM SIGCAS/SIGCHI Conference on Computing and Sustainable Societies}{August 16--19, 2023}{Cape Town, South Africa}
\acmBooktitle{ACM SIGCAS/SIGCHI Conference on Computing and Sustainable Societies (COMPASS '23), August 16--19, 2023, Cape Town, South Africa}
\acmPrice{15.00}
\acmDOI{10.1145/3588001.3609363}
\acmISBN{979-8-4007-0149-8/23/08}

\begin{document}

\title[Examining Indian Teachers' Sociotechnical Support Practices in Low-income Schools]{Pandemic, Hybrid Teaching \& Stress: Examining Indian Teachers' Sociotechnical Support Practices in Low-income Schools}

\author[A.Gavade]{Akanksha Y. Gavade}
\affiliation{
 \department{Computer Science}
 \institution{Lehigh University}
 \city{Bethlehem}
 \state{P.A.}
 \country{U.S.A.}
 }

\author[A.Sidotam]{Annie Sidotam}
\affiliation{
 \department{Education, Practice, and Society}
 \institution{University College London}
 \city{London}
 \state{}
 \country{U.K.}
 }
 
\author[R.Varanasi]{Rama Adithya Varanasi}
\affiliation{
 \department{Information Science}
 \institution{Cornell University}
 \city{New York}
 \state{N.Y.}
 \country{U.S.A.}
}


\begin{abstract}
Support plays a vital role in the teaching profession. A good support system can empower teachers to regulate their emotions and effectively manage stress while working in isolation. The COVID-19 pandemic has ushered in a hybrid form of education, necessitating the acquisition of new skills by teachers and compelling them to adapt to remote teaching. This new development further amplifies the sense of isolation prevalent amongst the teaching community. Against this backdrop, our study investigates the availability of sociotechnical support infrastructures for teachers in low-income schools while also looking into the support practices embraced by this class of teachers following the pandemic. Through 28 qualitative interviews involving teachers, management and personnel from support organizations, we demonstrate how teachers have largely taken the initiative to establish their own informal support networks in the absence of formal support infrastructures. Smartphones have significantly augmented these support practices, serving as both a valuable \textit{source} of support as well as a \textit{medium} for facilitating support practices. However, in comparison to other forms of support received from these sources, the availability of emotion-focused support for teachers have proven to be inadequate, creating imbalances in their support seeking practices. Our paper provides different contextual ways to reduce these imbalances and improve the occupational well-being of teachers.

\end{abstract}




\maketitle

\input{1-Introduction}

\input{2-Literature}

\input{3-Methods}
\input{5-1-Findings}
\input{5-2-Findings}

\input{6-Discussion}

\input{7-Conclusion}

\bibliographystyle{ACM-Reference-Format}
\bibliography{bibliography}

\end{document}

%% file: 1-Introduction.tex
\section{Introduction}

\begin{quote}
    \textit{``COVID-19 came and it forced us to use technology for teaching. No one asked us if we wanted to teach using technology. No one asked us if we could teach comfortably using technology \dots Now COVID-19 is gone but the technology still remains. I am yet to see someone care about us, reach out to us, and ask us how we are doing and not just ask us to implement another technology for teaching.''}
    -- \textbf{P09}, Government Teacher.
\end{quote}

The COVID-19 pandemic had a significant impact on essential workers worldwide, including individuals working in the teaching profession \cite{González_2021, Lancet_2020}. In a remarkably short period, teachers' work practices were disrupted and transformed. A key transformation was the extensive integration of technologies (e.g., smartphones, laptops) and the inculcation of remote-based learning  by schools \cite{ravi2022}. This transition has presented significant challenges for teachers, especially in low-income communities within the Global South, where technology-enabled work practices are just beginning to gain traction \cite{varanasi2019,cannanure2020}. Previously considered optional, these technologies have now become critical infrastructure for curriculum preparation, delivery, and other administrative tasks. The top-down integration of technology during the pandemic has further exacerbated the already demanding nature of the teaching profession, contributing to various forms of stress \cite{varanasi2021Investigatinga}. Understanding these stressors and mitigating them, especially for teachers in low-income communities within the context of hybrid technology-based teaching, is crucial for ensuring teachers' well-being and productivity in the classroom \cite{Zhao_Wang_Wu_Dong_2022}. 

Among other strategies, seeking support has proven to be effective for mitigating stress, thereby highlighting the the need for understanding how teachers in low-income communities engage in support-seeking practices \cite{varanasi2020}. By gaining insights into these practices we can identify potential solutions to alleviate stress and promote more efficient and sustainable teaching approaches in resource-constraint settings. However, most of the prior research on teacher-support has focused on educators in western contexts. There is extremely limited understanding around the sociotechnical support-seeking infrastructures available for the teachers in low-resource schools in the Global South. In order to fill this gap, we studied the support seeking practices of teachers within low-income schools within India.  

In particular, we answer the following research questions: \textbf{RQ-1} What kind of support do teachers seek in a post-pandemic educational setting? \textbf{RQ-2} What is the the role of technology in facilitating such support? To answer these questions we conducted a qualitative interview study with 28 participants from low-income private and public schools in India, comprising teachers, higher management, and personnel from support organizations assisting the schools (education-focused non-profits and companies). To examine the prevalent support practices manifesting in everyday work, we used a widely accepted framework within support literature that recognizes two types of support seeking practices \cite{Cutrona_Suhr_1992}. First, \textit{problem-focused} support practices which aim to regulate or alter a problematic situation that is contributing to the stressful situation \cite{lazarus1984stress}. Second, \textit{emotion-focused} support practices that are directed towards regulating emotional response to an often intangible issue \cite{folkman1980}.   

Our findings show that teachers lacked access to significant form of official support infrastructures to regulate the stress stemming from the transformed teaching practices. Instead, teachers relied on support infrastructures that they developed on their own to deal with new stressors. Smartphone technology played two key roles in this effort. Firstly, the smartphone acted as a \textit{source} of support for some teachers who were hesitant to reach out to their networks. Secondly, several teachers used smartphones as a \textit{medium} to receive and provide social support from colleagues at work. We also found asymmetries in the support practices of teachers which indicated that they were skewing heavily towards problem-focused methods. At the same time, teachers also faced challenges in receiving substantial emotion-focused support that could help them regulate their emotions in times of adversity. Our resultant discussion provides three critical ways to reduce the asymmetries in teachers' support seeking  practices, namely focusing on the lived experiences of the teachers, reducing power differentials, and improving the sustainability of support structures. Overall, our study contributes to the research of support in HCI context in the following manner:

\begin{enumerate}
    \item We explore two major ways in which smartphones are becoming integral for bottom-up support seeking infrastructures set up by teachers.
    \item We present asymmetries in these support infrastructures, contributing to skewed support practices.
    \item We propose three ways to reduce the asymmetries and improve occupational well-being of teachers.
\end{enumerate}

%% file: 2-Literature.tex
\section{Related Work}

\subsection{Support Practices: Coping Mechanism against Stress}
When individuals experience distress (negative stress, hereafter referred to as stress), they attempt to mitigate the effects through cognitive and affective actions \cite{lazarus20Stress}. Coping theory put forward by \citet{lazarus20Stress} recognizes such actions as \textit{coping} or \textit{support} mechanisms that are employed by individuals in response to the challenges in their well-being. Support mechanisms can manifest in different ways in our daily lives. They can take the form of individualistic responses such as disconnecting from the source of stress and voluntarily regulating one's emotions. They can also manifest socially when individuals rely on other people, groups, or organizations to deal with their stress. \citet{shumaker1984Theory} , in their seminal work, define this as social support theory, a process of ``exchanging resources'' between individuals where either the provider or the receiver has an explicit intention of improving the well-being of the receiver.
 
Three key factors play a significant role in the success of support mechanisms \cite{wills1991social,taylor2011social}. Firstly, the \textit{structural dimension} of support \cite{Chronister_2006}. It is determined by the individual's preference for seeking support through artifacts (e.g., smartphone, book), through their relationships with individuals (i.e., peer coping) or through organizations and institutions (i.e., institutional coping). 

Secondly, the \textit{function} of support in an individual's life \cite{house1981Work}. Support can be problem-focused, directed at ``managing or altering the problem causing the stress'' \cite{lazarus1984stress}. It can be \textit{informational}, sharing different types of inputs to help with stress-management. It can also be \textit{instrumental} or \textit{tangible}, providing individuals with explicit assistance to reduce stress. Support can also be emotion-focused, directed at ``regulating emotional response to the problem'' \cite{lazarus1984stress}. One such example is \textit{emotional} support that can provide an individual with emotional warmth and care to reduce their stress \cite{lazarus20Stress}. Other examples include \textit{network} support, creating a sense of belonging among people with similar interests \cite{Cutrona_Suhr_1992} and \textit{esteem} support, referring to expressions of ``regard for one's skills and abilities'' \cite{Cutrona_Suhr_1992}. Third factor is \textit{perceptual} which takes into consideration how an individual perceives and evaluates their social support \cite{sarason1991perceived}. Prior studies have established a strong positive relationship between social support and overall well-being \cite{turner1981social}. Multiple studies show that perceived social support by individuals helps in reducing stress and improving their well-being in personal \cite{house1988Social} and professional contexts \cite{Park2004}.

\bheading{Role of Technology in Social Support} 
Increasing development of internet and everyday technology has created new avenues for support through online-mediated communications in the form of emails \cite{Hutson_Cowie_2007}, forums \cite{Turner_Grube_Meyers_2001}, (micro-) blogs \cite{rains2011} and social networking sites \cite{Mustafa_Short_Fan_2015}. Early research on social support in online communication in the early internet-age has shown contradictory results, skepticism on one side \cite{kraut1998Social}, about whether increased internet use corresponds to reduced online connections and therefore reduced support, and other side demonstrating promising outcomes for social support \cite{shaw2002defense}. Recent research on ubiquitous devices like smartphones indicate them to be a crucial buffer for people. These technologies have the potential to increase an individual's social presence in online spheres thereby increasing their opportunities to avail support \cite{han2016}. However, there have been several inconsistencies in research about which kind of social support is really effective on online media; informational support \cite{Coursaris_Liu_2009}, emotional support \cite{Buis_2008}, or network support \cite{Ashley_Long_Rouner_Broadfoot_2011}.  

These inconsistencies have encouraged a narrower focus on individual elements in online spaces to understand their impact on social media \cite{SMOCK20112322}. A subset of research has focused exclusively on specific ecosystems within the broader internet medium. For instance, researchers have shown how certain applications (e.g., Snapchat) were associated with lower levels of social support than others \cite{Bayer_Ellison_Schoenebeck_Falk_2016}. Another set of studies has focused on more specific features and cues of online communities (e.g., likes, reactions, and emojis), known as paralinguistic digital affordances \cite{Hayes_Carr_Wohn_2016}, that provide social connections and support for individuals \cite{varanasi2018, Carr_Wohn_Hayes_2016}. Research has also focused on diverse actions associated with design features of technology and their affordances to understand how users feel supported. One such instance would be to look at how people perceived minor user actions \cite{Wright_2012}, like receiving a 'like' on their social media post, to be supportive \cite{Wohn_Carr_Hayes_2016}. Taken together, the overall research in this area indicates a need for further studies that establish a clear understanding of ubiquitous technology's role in seeking support.

\subsection{Support in Work}
Though we lack a clear understanding of ubiquitous technologies in the context of support, it is a fact that they have actively proliferated into individuals' work settings. Resultant forms of sociotechnical interactions through these technologies in work is contributing to stress in individuals, leading to new strands of HCI research to study and design technology-supported infrastructures to combat work stress. Some of these early HCI studies focused on applying established theoretical frameworks to measure the amount of support that employees perceived to be receiving in their work environments. Three such popular frameworks are the job demands-resource model that explores work resources of employees in response to job demands \cite{Bakker_Demerouti_2007}, conservation of resources theory that explores how workers use social support as a job resource \cite{halbesleben2010moderating}, and social exchange theory that explores how individuals reciprocate social support practices \cite{cropanzano2005social}. More recently, scholars in HCI have focused on examining support practices in the context of specific elements of work life like employee upskilling \cite{Dill2016}, performance of specific work functions \cite{yvette18, stdis15,broadbent2016intimacy} and finding new jobs \cite{radhika2020, Dillahunt21}. For instance, \citet{burke2013} in organizational work contexts found that weaker ties provide better support by bringing informational and instrumental support when individuals are coping with job loss.

Insights from these studies has motivated HCI researchers to study problems inherent in support infrastructures within under-resourced, emotional-labor-intensive job roles such as hospitality industry workers \cite{Huang_Van2018}, health care workers \cite{Tixier2009}, social workers \cite{Matthie2016}, gig workers \cite{Uttarapong_Bonifacio_2022} and teachers \cite{varanasi2019}. For instance, \citet{Poon2023} developed a computer-mediated intervention focused on peer-support for home care workers called ``sharing circles''. The intervention allowed workers to leverage storytelling practice to collectively reflect on their home care experiences and support each other. For the remaining of the section, we will focus on the teaching community around whom this study is situated.
    
\subsection{Support in Teaching Contexts}
Teaching is an emotionally demanding and labour-intensive job that requires teachers to spend a significant amount of time in silos while carrying out teaching responsibilities in limited resource environments \cite{hargreaves1994changing}. As a result, teachers experience stress and burnout, contributing to higher attrition rates \cite{Madigan_Kim_2021}. Technology-induced stress (or technostress) has become an additional source contributing to burnout due to the rapid integration of technology in teaching \cite{Zhao_Wang_Wu_Dong_2022}. Post COVID-19, the shift to hybrid form of teaching has made the teaching experience more isolated and stress inducing. For instance, \citet{González_2021} show how female teachers who did not have prior experience with technology experienced the most stress. Having an adequate support system in these contexts can be a crucial resource for teachers to improve not only their occupational well-being but also their professional development \cite{Richard2011}. Most early research in this area has focused on source, type, issues, and impacts of different types of social support. \citet{house1981Work}, in their work, pose this as a question of ``\textit{who} gives \textit{what} to \textit{whom} regarding \textit{which} issues''. For instance, one strand of research that studied peer interaction between teachers has argued that such practices, both offline and online, facilitated  informational, instrumental, and emotional support, thereby reducing the overall stress levels of teachers. \cite{dewert2003safe, kinman2011emotional, paulus2008can}. Another strand of research studied how traditional support practices (e.g., emotional, instrumental) in offline teaching translated to technology-enabled online peer interactions in the online forums \cite{clarke2014}. However, these comprehensive support seeking practices have been mostly shaped based on practices broadly resembling that of the Global North.

With proliferation of smartphones and cheap internet in Global South, teaching communities in low-resource settings have also started adopting technology in their profession \cite{cannanure2022, varanasi2020}. However, their technology-mediated teaching practices vary significantly when compared to the teachers in the Global North. Firstly, teachers in the Global South mainly rely on smartphones for teaching purposes, as compared to laptops and desktops in the Global North, due to their cheap cost and ease of adoption in resource-constrained teaching environments \cite{varanasi2019, cannanure2020}. Secondly, the smartphones used by teachers are often personal devices employed in work contexts, which add complexities that are not applicable to teachers teaching primarily through school-based technological infrastructure (e.g., smartboards, laptops) \cite{varanasi2020}. In addition, the pandemic pushed teachers, who had barely taught any classes remotely, to embrace hybrid teaching through smartphones \cite{ravi2022}. Early studies indicate how this push has contributed to reduced confidence, burnout, and attrition \cite{varanasi2021Investigatinga}. In this backdrop of changed landscape of teaching, it is essential to understand (1) the kind of sociotechnical support infrastructures that are prevalent and (2) the support practices that such infrastructures encourage teachers to take up in response to their stressors. Our work contributes to this research by exploring the support practices of teachers in low-income schools in post-pandemic settings. In particular, our work sheds light on the role of smartphones in facilitating the support activities.

%% file: 3-Methods.tex
\section{Methods}
To answer our research questions, we conducted a qualitative interview study spanning five months with diverse stakeholders in low-income schools. We describe the IRB-approved research in detail.

\input{4-table}

\subsection{Recruitment}
We recruited a total of 28 participants, including fourteen teachers, five higher management personnel from private and government schools that catered to low-income communities, and nine individuals from different education companies and non-profit organizations, which assist low-income schools in various capacities (hereafter collectively referred to as support organizations, see Table-\ref{tab:org-participants}). Following the self-selection sampling technique \cite{wainer2013drawing}, we advertised our study on WhatsApp groups, social media groups (e.g., Facebook), and offline social networks that were used by teachers and higher management from low-income schools. We also arranged informal information sessions with individuals who reached out with potential interest. We recruited final-study participants from among the individuals who signed up for our study. To be eligible for the interview, both teachers and higher management had to have been actively working in the last three months. To provide additional detail and context to the perspectives of teachers and higher management, we also recruited support organization personnel who worked in the participants' schools. To recruit personnel, we leveraged our long-standing networks and reached out to individuals who were either working or had recently worked in the participants' schools. Shortlisted organization personnel had worked for a minimum of three years (avg.$=6.1$ years, max.$=10$ years, S.D.$=2.6$ years) with the school faculty. 

\bheading{Procedure} Interviews were conducted in-person or remotely over phone, WhatsApp call, and Zoom, based on the interviewee's preference. A brief pre-interview call was set up to inquire about the participants' health, well-being, availability for a conversation and their preferred mode of communication. Participants were always provided with the option to reschedule or cancel the interview. The interview protocol was divided into three parts: we began the interview with high-level questions that acted as ice-breaking questions, encouraging the participants to share their current standing in their work, including their experiences and challenges in the last two months (e.g., \textit{``How has your work experience been over the last two months?''}). The next set of questions were intended to capture the different mechanisms that teachers employed daily to cope with the stress that they experienced at work, as well as to understand the immediate challenges they encountered around such practices (e.g., \textit{``Can you describe a recent example when you asked help from someone through the smartphone?''}). We also asked follow-up questions to better understand the support mechanisms employed by teachers (e.g., \textit{``Can you explain an instance where you had sought help from the management?''}). The final set of questions dealt with the different factors that influenced participants' support practices (e.g., \textit{``What issues did you experience from your peers on [WhatsApp] while asking for help?}''). For the interview sessions with higher management and support organization personnel, we extended this inquiry with a set of questions to triangulate the teachers' practices (e.g., ``\textit{What kind of medium did teachers prefer to receive support from you?} '') . In addition, we also inquired about their motivations and hesitations regarding the coping mechanisms employed by the teachers. All the questions were open-ended, free of technical jargon, and designed to be neutral rather than leading. Interviews were conducted in Hindi, Telugu, or Marathi by the authors.

\bheading{Data Collection \& Analysis}
Our data consisted of 32 hours of audio-recorded interviews and 45 pages of field notes. All the interviews were translated into English, transcribed and analyzed using thematic analysis in MAXQDA. The process of analysis involved all authors taking several passes on the transcribed interviews to familiarize themselves with the participants' experience and internalize their narratives. All authors conducted open-coding on the transcription without any preconceived theoretical assumptions. The authors established credibility by engaging with the data over several weeks. Significant disagreements between authors during the process were resolved through multiple rounds of peer-debriefing \cite{Creswell2000}. The authors chose the final codes carefully by merging the overlapping codes. The final codebook consisted of 54 codes. Example codes included \textit{``informal agreements''}, \textit{``trust issue''}, and \textit{``support temporality''}. The codes were further mapped and categorized into appropriate themes. Example themes included \textit{``support apprehensions''}, \textit{``support layers''}, and \textit{``dual nature of support''}.

\bheading{Ethical Considerations}
Our overall interpretations in the paper are shaped by our education and research experience. While all authors have received education in western contexts, they also have significant experience teaching in India, with two authors spending over four years teaching in these settings. All the authors are of Indian origin. All but one are women. The interviews were conducted by male and female authors to balance individual biases and capture diverse perspectives.

%% file: 4-table.tex

 

\begin{table}
 \center
 \renewcommand\arraystretch{1.3}
 \footnotesize
 \begin{tabular}[t]{|p{.75in}|p{2.35in}|}
 \hline
 \multicolumn{2}{|l|}{{\bf Teachers} (n=14)} \\ 
 \hline
  Gender & 
       \begin{tabular}{ll}
           Women: 8 & Men: 4\\
       \end{tabular}\\
\hline
 Age (years) & 
       \begin{tabular}{llll}
         Min: 22 & Max: 43 & Avg: 34.5 & S.D: 8.9\\
       \end{tabular}\\
\hline
Education (degree)  & 
       \begin{tabular}{lll}
           Bachelor's: 6  &  Master's: 8\\
       \end{tabular}\\
\hline
Region  & 
       \begin{tabular}{p{.49in}p{.48in}p{.58in}p{.45in}}
           Karnataka: 3 & Telangana: 6 & Maharashtra: 3 & Kerala: 2 \\
       \end{tabular}\\
\hline
Experience (years) & 
       \begin{tabular}{llll}
           Min: 4  & Max: 25  & Avg: 11.64 & S.D: 6.9 \\
       \end{tabular}\\

\hline
Phone use (years) & 
       \begin{tabular}{llll}
           Min: 2  & Max: 5.5  & Avg: 3.8 & S.D: 1.2 \\
       \end{tabular}\\
\hline
 Focus Subject & 
     \begin{tabular}{llll}
           Languages: 6 & Science: 8 & Math: 6 & Social Studies: 8 \\
       \end{tabular}\\
\hline
School Type & 
     \begin{tabular}{ll}
           Government: 6 & Private: 8 \\
       \end{tabular}\\
\hline
\hline
 \hline
  \multicolumn{2}{|l|}{{\bf Management} (n=5)} \\ 
 \hline
 Gender & 
       \begin{tabular}{ll}
           Women: 4 & Men: 1\\
       \end{tabular}\\
\hline
 Age (years) & 
       \begin{tabular}{lll}
           Min: 46 & Max: 53 & Avg: 2.9\\
       \end{tabular}\\
\hline
 Roles & 
     \begin{tabular}{ll}
       Management Board: 2 & Principal: 3 \\
     \end{tabular}\\
\hline
Experience (years) & 
       \begin{tabular}{llll}
           Min: 10 & Max:25 & Avg: 17.15 & S.D: 7.5 \\
       \end{tabular}\\
\hline
School Type  & 
       \begin{tabular}{ll}
           Government: 3 & Private: 2 \\
       \end{tabular}\\
\hline
 Region  & 
      \begin{tabular}{llll}
           Karnataka: 1 & Telangana: 2 & Delhi: 1 & Kerala: 1 \\
       \end{tabular}\\
\hline
 \hline
 \hline
  \multicolumn{2}{|l|}{{\bf Support Organization Personnel} (n=9)} \\ 
 \hline
  Gender & 
       \begin{tabular}{ll}
           Women: 6 & Men: 3\\
       \end{tabular}\\
\hline
 Age (years) & 
       \begin{tabular}{llll}
           Min: 25 & Max: 36 & Avg: 29.3 & S.D.: 4.1\\
       \end{tabular}\\
\hline
Experience (years) & 
       \begin{tabular}{llll}
           Min: 3 & Max: 10 & Avg: 6.1 & S.D.: 2.6\\
       \end{tabular}\\
\hline
Role & 
       \begin{tabular}{llll}
           Co-founder: 3 & Manager: 3 & Lead: 3\\
       \end{tabular}\\
\hline
Education (degree) & 
       
          \begin{tabular}{ll}
           Bachelor's: 3  &  Master's: 6 \\
       \end{tabular}\\
      
\hline
Organization & 
       
          \begin{tabular}{p{2.3in}}
           321 (2), Godrej foundation (2), Meghshala (2), Mantra4Change (1), Alokit (1), Simple Education Foundation (1)\\
         \end{tabular}\\
      
\hline
\end{tabular}
\caption{Demographic Details of Teachers, management, and support organization personnel}
 \label{tab:org-participants}
 \vspace{-6mm}
\end{table}

%% file: 5-1-Findings.tex
\section{Findings}

Our overall findings  reveal significant inconsistencies across different types of support available to teachers. The majority of the participants (n=22) reported a lack of any formal support, whether remote or in-person, both before and after the pandemic. Instead, teachers were expected to develop and rely on their own informal support structures. This outlook stemmed from the managements' expectation that teachers should be resilient and self-reliant. Several principals and school board members used words like ``selfless'' and ``community-focused'' to characterize the teaching profession,   while also placing the responsibility on teachers to reach out and seek the needed support. As \fixme{P12}, a government school principal pointed out, ``\textit{Teachers know when they need help. We are always available. If they have any issues, they can always reach out to us}''. Due to the lack of explicit support infrastructures, we observed a great variability in support activities and the associated role of smartphones in them. In the following sections, we present two major ways in which teachers utilized smartphones to develop their own bottom-up support practices.   


\subsection{Seeking Support in Smartphone Apps: A Sign of Self-reliance \& Strength}

A fraction of teachers nurtured their own support infrastructures by relying on smartphones as a \textit{source} of support. For a subset of teachers (n=3), this motivation stemmed from their hesitation to reach out for social support from their professional network. They perceived reaching out for support as a sign of imperfection or weakness in an otherwise idealistic job -- a job that is expected to provide selfless emotional labor. These teachers conformed to this ideal image by solely \textit{providing} assistance instead of \textit{asking} for it from their colleagues, management, or family members. A few teachers also feared that seeking support could be perceived by the management as a sign of incompetency. For example, \fixme{P25}, a Math teacher at a low-income private school with a teaching experience of over twelve years, spoke of the challenge of adapting to the changing curriculum post-pandemic. Despite facing difficulties with the new curriculum, he hesitated to seek support as he believed it might undermine his expertise gained over the years. Instead, he focused on finding solutions on his own by watching YouTube videos of others teaching the same curriculum. Another teacher \fixme{P18}, teaching English in a government school, shared how she used smartphone as a source of informational support and a mode of distraction: 

\begin{quote}
    \textit{``I can’t take support from others as many people depend on me for all kinds of support \dots they think I am strong because I give good advice and help them. So, I don’t feel like showing them my vulnerable side. Most of the times, I try to handle my problems myself \dots In school or at home, I sit in a corner and watch video on how to teach difficult poems to kids. Sometimes, if I have a lot of stress, I have an app to write daily diary, but it's not possible every time. Then I just listen to music or read Shayari (Urdu poetry) on phone. Expressing myself by writing status on WhatsApp gives me great relief too.''}
\end{quote}


Similarly, interviewees shared how the onset of COVID-19 pushed teachers to work in isolation while trying to understand and incorporate new pedagogical norms. They were also taking on the additional workload of teachers who were leaving their jobs. The resultant pressure left them too exhausted to make the emotional and physical effort to seek out social support. Moreover, teachers shared how they realized that their peers were in a similar situation, fearing rejection if they reached out for any form of support. Post COVID-19, teachers experienced similar challenges when schools reopened. They had to attend to critical issues like helping students recover from longitudinal learning gaps that emerged during the pandemic, in addition to embracing new forms of hybrid education. It meant that teachers had to forfeit their time for socialization, like lunch- and tea-breaks, to take up full-day teaching schedules, leaving fewer opportunities for social support. 

\subsubsection{Problem-focused Support through Smartphones}

The reasons outlined above made it easier for a subset of teachers to seek support directly from smartphones instead of reaching out to their social networks. In particular, several teachers relied on their smartphones for problem-focused support. To cope with the pressure of unanticipated pedagogical changes, such as introduction of remote teaching, several teachers relied on Google Answers and YouTube videos to find solutions to their problems with technology-mediated teaching practices. For instance, \fixme{P28}, a Science teacher in a low-income private school used YouTube videos in her local language to receive \textit{informational} and \textit{instrumental} support and to find solutions for technological issues that she was facing while teaching an online class:
\begin{quote}
    \textit{``I don't ask for help\dots I don't think the other teachers can help me. They are struggling themselves\dots During COVID-19, I had issues playing video in Zoom, there was no audio. Google and YouTube were the only place for solutions when I was breaking my head over the problem. YouTube is effective because it is like someone is telling me where to click without me having to be burden to someone else\dots The downside is that it is either a hit or it is a miss which consumes a lot of time''}
\end{quote}

Despite the significant time it took to find solutions from the videos, the accessibility of smartphone-based interactions made teachers like \fixme{P28} feel autonomous and self-sufficient. The assistance was similar to being helped by another peer, without having to feel that they were being a "burden" on colleagues and family. Such context-based online resources also allowed teachers to not feel privy to the judgment and potential gossip between their peers, that would create further stress. 
\begin{figure*}[t]
   \centering
  \includegraphics[width=\textwidth]{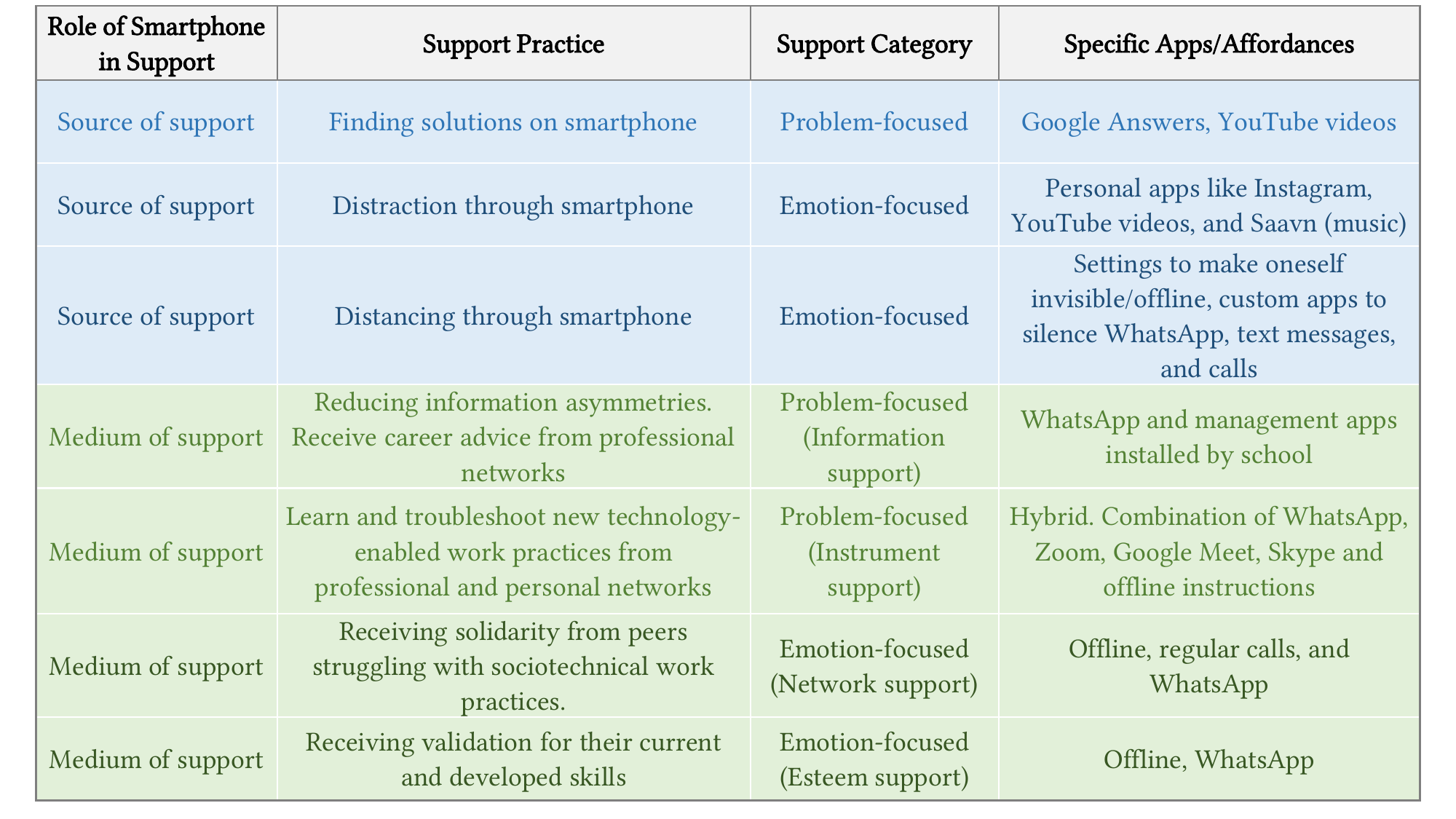}
  \caption{Table summarizing key support seeking practices of Indian teachers in low-income private and government schools . }
  \label{summary table}
 \end{figure*} 

\subsubsection{Emotion-focused Support through Smartphones}
Several teachers also utilized smartphones for emotion-focused support. One commonly used strategy by teachers for emotion-focused support was technological \textit{distraction}. When teachers found an issue overwhelming and stressful, they distracted themselves from their negative stressors for brief periods of time by engaging in digital activities on their smartphones. These distractions included switching to personal tasks such as browsing Instagram, watching YouTube videos, or listening to songs. 

Another emotion-focused strategy was the usage of smartphones to \textit{distance} themselves from the person or the event contributing to their stress. A common strategy included changing indicators that made them invisible online while working from home on apps like Zoom, WhatsApp, Google Meet, and Microsoft Teams. It helped them avoid stressful engagements with peers and management by distancing themselves in the online space while continuing their own work. In extreme cases, strategies were also aimed at completely disconnecting themselves from work through custom apps designed to silence WhatsApp, text messages, and call notifications from the management. \fixme{P14}, the principal of a government school, shared how teachers in her school used such mechanisms frequently to disconnect from troublesome parents who negatively impacted their work lives and made them feel ``helpless''.


%% file: 5-2-Findings.tex
\subsection{Seeking Support From Social Connections: Smartphone as a Medium}

While a few teachers used smartphones as a medium for self-reliance, most teachers (n=11) relied upon smartphones to be a \textit{medium} through which they could build their informal support infrastructures and seek social support from their professional (peers, management and support organizations) and personal (family members and friends) connections. While teachers already had the option of seeking support from their networks through smartphones prior to COVID-19, it was only after the onset of the pandemic that they embraced the option completely, given the challenge of remote learning. For these informal infrastructures to be successful, the teachers who were seeking support and the individuals who were providing support had to rely on common cues that both the former and the latter understood. The first category of cues were subtle changes in work-related attitudes or behavior of an individual, including lack of attention, increased temperament, careless mistakes and/or nonchalant attitude towards their students. For example, when a peer who is otherwise punctual and meticulous, began sharing irregular attendance sheets, made silly calculation errors, and seemed distracted on calls during lunch breaks, \fixme{P19}, a Hindi language teacher in a government school, saw these actions as cues that prompted her to reach out to her peer to provide assistance.

The second category of cues included smartphone-mediated actions that indicated signs of stress. These cues included unusual WhatsApp and Facebook status changes, abnormal profile pictures updates, melancholic forwards, repeated deactivation and reactivation of profiles on social media, leaving work WhatsApp groups, and being offline for lengthy periods of times. \fixme{P09}, a veteran government Science teacher shared how she became concerned about her peer's well-being and reached out to offer support when her peer suddenly removed her husband's photo from her WhatsApp profile picture and regularly went offline for long periods of time.
 
\subsubsection{Problem-focused Support from Peers through Smartphones}
Teachers frequently sought problem-focused support from their networks, utilizing smartphones as the primary medium. These behaviors arose as a means to address the overall work stress stemming from various issues, particularly the shift to online teaching during the onset of the COVID-19 pandemic. Within this support category, we observed a few distinct patterns of support-seeking behavior among teachers. 

\paragraph{Information Support}
One of the biggest challenges during the COVID-19 pandemic was the information asymmetry experienced by teachers working in low-income private schools in comparison to government and high-income schools. Critical information, required for teachers to fulfill their responsibilities, come from the management headquarters that oversee the functioning of the schools and their curriculum. These include school schedules (e.g., ``\textit{Is the government going to open the schools or not?}''), examination schedules, guidelines, assessment instructions, attendance weightage, extracurricular requirements, and professional development opportunities. This information flowed through many top-down channels before reaching the low-income schools. While teachers in government schools had strong information channels (e.g., state-based WhatsApp groups \cite{varanasi2021Tag}) that helped them get guidance during and after the COVID-19 pandemic, most teachers in low-income private schools had to rely on information support from different sources. One way to fill their information gap was to establish connections and seek support from teachers in governments schools who received the information faster. \fixme{P16}, an English teacher in a low-income school shared:

\begin{quote}
    \textit{``We have [name omitted] Ma'am, who has worked in different schools before. She has excellent network with the teachers from other schools. When we were all struggling in COVID-19, she got the syllabus and exams schedule information very fast from the government school teachers on WhatsApp. Problem is that we are a low-income school \dots By the time our principal gets information and shares with us, we have very less time to do anything and that puts pressure on us \dots She also later introduced me to one of the teachers in the government school \dots now I have my own network. I ask that new teacher.”
''}
\end{quote}

Information support extended beyond managing information asymmetries to cope up with managements' increased demands and overall loss of teacher workforce. The management introduced several disparate tools that were supposed to act as a `unified learning management system'. For example, one combination that a few schools tried was using Zoom for teaching, Google Forms for capturing attendance and conducting exams, and WhatsApp for coordinating corrections and classroom management. Teachers found it extremely stressful to adjust with several new tools and online pedagogical practices at the same time. As a result, they relied on personnel from support organizations who were in touch with them during the COVID-19 pandemic. Even though the support personnel were not present in-person, they provided much required information support to fill the gaps in teachers' mental models on how to use the tools. 


To manage the unrealistic work expectations from managements, teachers also requested their peers to take up the role of information intermediaries to share their own online work load. \fixme{P11}, a Social Science teacher in a government school, shared how he sought information support from his peer to attend administrative meetings with the management during an extremely busy work day, while he combined his class with his peer's class and taught everyone together. The teachers sought and acted on the meeting information that they received from their peers later. The unpredictability of the COVID-19 pandemic, beyond work life, also extended teachers' information support seeking practices with their peers and support organizational personnel to their personal lives. One of the most common modes of information support sought by teachers working in affordable private schools was career advice as these schools were drastically reducing teacher workforce or even closing down altogether due to lack of fee payment by the parents. Several teachers reached out to support personnel for information, seeking leads on teaching opportunities, strategies to upskill themselves, and feedback on their application material.

\paragraph{Instrumental Support}
The post-COVID-19 hybrid teaching practices also put significant pressure on the teachers' work lives. Consequently, several teachers relied on instrumental support in the preparation phase of their work to learn and troubleshoot the new technology-based work practices, including figuring out how to conduct classroom management activities such as disciplining elementary school students through screens, coordinating structured responses, encouraging participation, and seeking support of parents in the process. Teaching through technology required them to design learning materials for remote teaching and make formative assessments. For example,  \fixme{P15}, a new Mathematics teacher who joined a low-income private school during the pandemic, constantly struggled with taking online attendance and managing a classroom of thirty children on Zoom. P15 shared how she took the support of their peers, who hopped on to WhatsApp or a Zoom call to help her take attendance while she managed the classroom and focused on teaching. The teachers used this information to follow-up with the absent students. P\fixme{27}, who helped teachers similar to P15 shared:

\begin{quote}
    \textit{``I also support her [a teacher] through virtual mode. She does not know much about computers and technology \dots but she reaches out to me and messages me about making a PowerPoint presentation and Google Forms. I made it and sent it to her WhatsApp and she used it in her classroom. If we have to enter the marks, admin staff from school guides us. We just follow the same process. If we are not in school, then we take the help of a peer or our brother or sister and do it.''}
\end{quote}

Unlike informational support, we found that teachers sought instrumental support from individuals in both professional and personal networks. For example, teachers took help of their friends and family members for support around technological features and issues of smartphones while they sought support from their colleagues on how to apply the technology to pedagogical practices. Common examples in this category included, ``\textit{how to present materials to the classroom}'', \textit{``how to manage young students''}, \textit{``how to plan and execute online in-class activities''} remotely. Balancing instrumental support between professional and personal connections helped teachers deal with the increased technostress during the height of the pandemic. In extreme situations, COVID-19 pushed several teachers' families into health and economic volatility compelling teachers to seek financial assistance from colleagues working in governments schools who had relatively stable jobs.  

While teachers successfully reached out for information support both before and during the COVID-19 pandemic, teachers faced substantial setback in receiving instrumental support during the pandemic. This was because instrumental support often had to be sought remotely through smartphones and often the support was for learning how to use smartphone for teaching, creating an additional layer of complexity. For example, teacher \fixme{P18
} described what the instrumental support for teachers before the COVID-19 pandemic entailed:

\begin{quote}
    \textit{``I used to stay with the new teacher during their initial classes, introduced the new teacher to students and sat there along with the teacher throughout the first class to build the trust and bond between them. \dots Post class, I used to give them feedback about their teaching style and their learning materials and provide assistance when they were teaching by managing the classroom, like motivating children to answer the questions. These small but important practices have become extremely difficult to carry out after COVID-19. ''}
\end{quote}

In-person instrumental support was more effective as the focus of the support was entirely on providing assistance around teaching and classroom management. The same teacher explained how her instrumental support practices shifted to technology troubleshooting and low-level coordination activities that made providing support less rewarding,

\begin{quote}
    \textit{``The help has shifted to telling the teacher that the child has muted the mic, you have to call the child's name, ask them to unmute. If a student who is not part of the subject joins your class by mistake, then how to remove the child. I have to show them two or three times before they can start doing themselves \dots help is more about about non-teaching stuff.''}
\end{quote}

To overcome this challenge, many teachers began providing a hybrid form of instrumental support for technology-enabled teaching. Teachers initiated this support by sending informational resources about the topic in question through WhatsApp or email. Once the teacher reviewed the information resource, they shaped their understanding by providing contextual information over a call or in-person meeting. For example, when a support organization personnel, \fixme{P01}, saw that an English teacher was struggling to teach the concept of simile and was stressed because of the pressure from the principal, she suggested teaching resources that could help her learn new techniques. She then called her and then walked her through example classroom scenarios and ways implement those resources. 

\subsubsection{Emotion-focused Support from Peers through Smartphones}
Teachers' informal support infrastructures were skewed towards receiving and providing problem-focused support. Compared to these practices, we found limited development of emotion-focused support practices in both offline and online mediums. A key hesitation for the teachers in seeking emotion-focused support was the lack of confidentiality and nurturing environment that the workplace provided. In the beginning of the pandemic, top-down hierarchies promoted teacher productivity to improve student outcomes, often at the cost of their own personal space. Teachers were encouraged to compete with their peers to demonstrate improved outcomes in their subjects.  Surveillance structures, such as ``teacher monitors'', were established to ensure that teachers were performing up to their potential without any distractions. Teachers and support organization personnel shared how these practices created a culture of gossip and rumors that brought a sense of apprehension and curtailed teachers from sharing problems that were emotion-focused with their peers. \citet{varanasi2021Investigatinga}, in their recent study, showed similar top-down structures that didn't have any provision of safe spaces, contributing to emotional stress amongst teachers and pushing them to engage in emotion-focused support practices. New hybrid education mediums moved these competitive structures and surveillance practices online during and after the COVID-19 pandemic, increasing apprehensions to seek support. 

Support organizations, that reached out to different schools to assist teachers in transitioning to hybrid teaching during the pandemic, had to put in a lot of effort to help teachers overcome their fear of sharing their emotional experiences in anticipation of retaliation from management and peers. Support person \fixme{P06}, working in a start-up focused on overall school capacity development, shared the required effort in her own words: 

\begin{quote}
    \textit{``I really had to convince teachers that I was on their team and I wasn't a spy for school management. They are definitely  wary of the school management because the management also tends to treat these teachers like kids. They scold the teachers and say, `this is not done right' or `sit  straight in a zoom call' \dots  COVID-19 has brought out a weird way of disciplining teachers in the school \dots But when the teachers realized that the conversations that we were having were indeed confidential that I haven't gone and complained, which meant no repercussions for them. So being patient really helped me build that trust.''}
\end{quote}
Unlike problem-focused support, emotion-focused support practices required teachers to establish profound trust and foster strong bonds before seeking assistance. Even after cultivating the necessary trust, teachers adhered to distinct practices that set emotion-focused support practices apart from other forms of support. For instance, some teachers favored seeking support offline or via voice calls instead of text messages, apprehensive that written communication could be used as evidence and potentially lead to repercussions from management.

\paragraph{Network \& Esteem Support}
Within the limited instances of emotion-focused support practices that the teachers shared, network support was quite prominent \fixme{(n=4)}. Network support is a type of emotion-focused support that focuses on providing oneself with a sense of belonging with individuals with similar interests or commonalities \cite{Cutrona_Suhr_1992}. Teachers who were uncomfortable or new to technology-mediated teaching received assistance from support organization personnel in forming offline groups. They achieved this by calling teachers in their neighborhood districts to find other teachers who were struggling to share their challenges. While these groups consisted of teachers who didn't know each other previously, the groups provided a sense of solidarity and encouragement to come together and talk about the challenges that they were encountering in their preparation and teaching process. \fixme{P21}, who taught Telugu language in a low-income private school, shared how she, along with seven more teachers in other schools across the district, curated and maintained a contact list of senior teachers who joined their calls and provided the required emotional support to new teachers joining the network. Teachers who were struggling with the administrative aspects of their jobs, such as not receiving regular salaries in the early periods of the COVID-19 lockdown, also formed common WhatsApp groups to find a sense of togetherness and reduce anxiety around the lack of information and clarity in their professional lives.

In addition, support organizations personnel also provided the much needed esteem support for teachers, validating the teachers' skills and bolstering their intrinsic sense of purpose \cite{Cutrona_Suhr_1992}. Examples of this mainly included supporting teachers in instances of failure or when they made mistakes. Support personnel shared how school managements rarely promoted learning-by-failure \cite{Bartholomew_Mentzer_2022, varanasimaking}, a key concept in education where teachers are provided a safe space to learn through failing and to grow from their mistakes. Support organizations gave the appropriate space for teachers to reach out in their moments of failures and receive validation of their teaching skills. Common situations included when teachers struggled to teach a post-pandemic classroom with large variability in learning skills or lost confidence in their own pedagogical and content knowledge. \fixme{P28} shared how a personnel from a support organization used to ``encourage'' her remote teaching ideas and skills even though she was struggling with it. Another place where teachers used to receive their esteem support from was over impromptu online meetings with teachers from other schools whom they had not met in a long time. Teachers shared how such unexpected encounters initiated formal conversations that were focused on solidarity, where teachers came together to bolster their own confidence in their teaching skills and to encourage each other to keep on working.  





%% file: 6-Discussion.tex
\section{Discussion}

The lack of formal support infrastructures for low-income teachers in HCI4D settings is a cause for concern, as indicated by our research findings. Prior studies have highlighted the positive impact of a proper support system on teachers' overall stress management capability, particularly in relation to technology-induced stress \cite{varanasi2021Investigatinga}. The absence of formal support further hinders teachers' ability to effectively alleviate stress. This becomes particularly challenging in a post-pandemic setting within India, where teachers are grappling with the demands of hybrid teaching. In this challenging context, teachers' own grassroot level informal support networks helped regulate some of these stressors. However, these networks have their limitations and cannot fully address the issue. For instance, teachers tend to be hesitant in seeking emotion-focused support, resulting in imbalanced support structures. In this section, we present our initial recommendations to bridge this gap and improve teachers’ occupational well-being. To achieve this, we seek inspiration from \citet{pendse2022}’s framework to root our recommendations in the local contexts of teachers. \citet{pendse2022} define their framework for localized technology-mediated forms of care where sociotechnical practices are grounded in the lived experiences of individuals who need care. It is focused on reducing power differentials and promoting structures that are sustainable by being considerate of community contexts and constraints.

\subsection{Tapping Lived Experiences: Exposing Teachers' Collective Narratives}
A key reason for the absence of official support structures from the management is the lack of a clear visibility of teachers' issues from their perspective. With hybrid work, activities are often hidden from the direct purview of the employer \cite{Bowker_2016}. This also means that the stressors arising from work are also invisible, leading to an inaccurate impression that the employees do not need support. To rectify this situation and provide better support, it is essential to increase the visibility of the \textit{lived experiences} \cite{pendse2022}. 
A common solution to this issue in the western-work contexts has been the implementation of monitoring-based solutions to increase worker visibility. However, such solutions are often misused by the management to enforce status quo, giving rise to the risk of surveillance practices that can in turn create new sources of stress \cite{liu2017whispers, marathe2016ict}. An alternative approach involves bringing to light the collective narratives of teachers’ lived experiences, allowing for a more nuanced understanding of their challenges while avoiding the pitfalls of surveillance. For instance, collecting and disseminating oral histories at school or district level can expose issues that were not previously visible to the managements’ attention. Such initiatives can also challenge the structural assumptions that teaching roles are inherently noble and selfless which push teachers to disregard any kind of hardships that arise in their work. This shift in perspective is essential to foster a supportive and empowering environment for teachers.

\subsection{Reducing Power-differentials: Support Organizations as Mediators of Trust}
To develop collective narratives, it is essential for teachers to trust their peers, higher management, and the overall work ecosystem. Our findings indicate a sense of eroding trust on the side of teachers. In particular, teachers found their peers to be less \textit{reliable} and the management to lack \textit{openness} to seek different forms of emotion-focused support. Cultivating these qualities is crucial for creating a high-trust environment in schools  \cite{handford2013teachers}. But in reality, we saw how informal but critical support structures that required high levels of trust, were not utilized as much. To fill this gap, personnel from support organizations can serve as invaluable resources. By operating outside the traditional hierarchical structures of the school ecosystem \cite{varanasi2020}, they occupy a neutral position in schools, enabling them to provide support in unconventional ways. For example, support personnel can have more flexibility than the management to engage in meaningful conversations with the teachers outside of regular school hours, where teachers may have more time and feel more comfortable seeking emotional support.  Furthermore, support personnel can act as crucial intermediaries between teachers and management, as well as peers, by offering anonymous and collective feedback on the challenges teachers face. This process can reduce burden on the management, allowing them to provide support for certain issues (e.g., problem-focused support), complementing the support that organization personnel can provide (e.g., emotion-focused support)

\subsection{Sustainable Sociotechnical Support Structures: Hybrid Solutions for Hybrid Future}
In this final section, we present recommendations for researchers and designers who can play an active role in developing sociotechnical interventions and can better assist teachers’ emotion-focused support requirements. Popular technology-based interventions (e.g., chatbot assistance \cite{Fitzpatrick2017}) that work well to provide support in western contexts do not work in resource-constrained settings. Moreover, these interventions can add to the cognitive stress experienced by teachers who are already grappling with the challenges of technology-mediated hybrid teaching. To address these concerns, previous researches have emphasized the importance of carefully considering technologies in such contexts to mitigate technostress and burnout among teachers \cite{varanasi2021Investigatinga}. Building on this line of research, we propose the notion of \textit{collective hybrid support} interventions that balance subject matter expertise and the localized contexts to deliver effective support. This approach involves multiple actors working in synergy to deliver support. For instance, in the case of emotion-focused support, collective hybrid support interventions can bring together teacher experts who have extensive experience in handling emotion-focused issues but lack local contexts and support personnel who posses teachers’ trust and are aware of the local contexts and norms (e.g., cultural taboos associated with mental health) but lack the expertise in mental health to augment each other and provide holistic mental health support for teachers.

%% file: 7-Conclusion.tex
\section{Limitations and Conclusion}
Overall, our study provides exploratory insights into how teachers in low-resource settings establish and nurture their sociotechnical support infrastructures in post-pandemic work settings. In the absence of top-down support infrastructures,  teachers often took it upon themselves to establish their own practices. Smartphones played a key role in these practices by acting both as a \textit{source} of support as well as a \textit{medium} to connect teachers with their networks to seek social support. Out of all types of support, teachers struggled the most to find adequate emotion-focused support as compared to problem-focused support, presenting an excruciating need for sociotechnical interventions that can integrate support organizations to provide contextual emotion-focused support. We end our paper by providing key recommendations to fulfill this need. Our work has usual limitations posed by qualitative research such as the sample size being small. Also, the primary focus of our study being the experiences of teachers in India, limits its generalizability beyond the Global South. 

\section*{Acknowledgments}
We are grateful to our participants who provided valuable time and experiences for this study. We thank Nitya Agarwala who contributed to this research in its early stages. We also appreciate the guidance and feedback provided by Dr. Nicola Dell to make this study a success. Finally, we also thank the external reviewers for value feedback on the paper.